# CT Saturation Detection and Compensation: A Hybrid Physical Model- and Data-Driven Method

*Songhao Yang, Member, IEEE, Yubo Zhang, student member, IEEE, Zhiguo Hao, Member, IEEE, Zexuan Lin, Baohui Zhang, Fellow, IEEE*

*Abstract*—Current transformer (CT) saturation is one of the dominant causes of relay protection devices' malfunctions, which pose a threat to the safe operation of the power system. To address this problem, we propose a hybrid physical model- and data-driven method. The method firstly detects the CT saturation and then compensates it to reproduce the real waveform. Considering the multi-factor and strong nonlinearity of CT saturation, a data-driven model, namely the Fully Convolutional Network (FCN), is built to detect the operation status of CT. As for the compensation, a physical model of short-circuit current is used for its conciseness and universality. Through tactfully integrating the data model and the physical model, the proposed method is endowed with two major merits: the arduous adjustment of universal thresholds and parameters in existing methods is avoided, and the deficiency in generalization and interpretability of the data-driven method is assuaged. Simulation and experimental results verify the effectiveness of the proposed method. Furthermore, its application potential to future protection is explored.

## Nomenclature

*Acronyms*

| | |
|---|---|
| CT | Current transformer. |
| EKF | Extended Kalman filtering. |
| FCN | Full convolutional network. |
| NLS | Nonlinear least squares. |
| LM-NLS | Levenberg-Marquardt nonlinear least squares. |

*Variables and Parameters*

| | |
|---|---|
| $i_1$ | Short-circuit current/Primary current of CT. |
| $i_2$ | Secondary current of CT. |
| $i_m$ | Magnetizing current of CT. |
| $i_2^*$ | Normalized output current of CT. |
| $I_m$ | Amplitude of the steady-state short-circuit current |
| $\theta$ | Initial phase angle of fault. |
| $T_1$ | System time constant. |
| $\omega$ | System angular speed. |
| $f$ | System frequency. |
| $T_2$ | Time constant of the secondary closed-loop circuit of CT. |
| $\phi_r$ | Remanence of CT. |
| $S$ | Serial saturation information. |

## I. Introduction

CURRENT transformer (CT) is a kind of essential measuring device for protection and state monitoring in the power system, of which function is to accurately measure the current of the primary devices. However, the accuracy of CT may be affected by CT saturation which usually occurs when the short-circuit current especially its DC component is enormous. Serious CT saturation will cause the malfunctions of the relay protection devices, posing a threat to the safe operation of the power system. Therefore, how to deal with the issue of CT saturation has attracted widespread attention.

As magnetic saturation is an inherent property of ferromagnetic materials, a straightforward solution to CT saturation is to design and manufacture the non-ferromagnetic CTs, e.g., optical current sensor [1] and Rogowski coil [2]. However, such transducers suffer the problems of noise sensitivity and accuracy, thus they are not widely used in practice. For ferromagnetic CTs, introducing an air gap in the core magnetic path can significantly decrease the remanence and keep CT unsaturated [3]. TPY-type CTs [4] and TPZ-type CTs [5] are designed according to such technical ideas. The TPY-CT has a small air gap in the core and its remanence is less than 10% of the saturation flux. Due to its excellent transient performance, TPY-type CT is widely used in relay protection especially in high-voltage transmission power systems. Moreover, TPZ-type CT eliminates remanence by increasing the air gap. However, the poor accuracy of TPZ-CT impedes its application. In addition, the air gap of TPY-CT and TPZ-CT also lead to a larger size and higher cost. Benefited from the stable performance and acceptable cost, the traditional CTs, i.e. the closed-core P class CTs, are still widely used in current power systems especially in distribution systems and non-high voltage transmission systems.

Retaining the ferromagnetic core closed, scholars attempt to keep the CT unsaturated by changing the circuit structure. Power electronic-based method [6], [7], and voltage feedback-based method [8] were proved to be robust and efficient. However, these methods need to reconstruct the hardware structure of CT, which leads to high costs. By contrast, the software-based methods are cost-effective, which can be divided into three types: 1) CT model-based methods; 2)

This work was partially supported by the Joint Funds of the National Natural Science Foundation of China and State Grid Corporation of China(U1866603), National Natural Science Foundation of China (52007143), China Postdoctoral Science Foundation (2021M692526) and Open Fund of State Key Laboratory of Operation and Control of Renewable Energy & Storage Systems (China Electric Power Research Institute).

All authors are with Shaanxi Key Laboratory of Smart Grid, Xi'an Jiaotong University, Xi'an,China(e-mail:{songhaoyang, zhghao, bhzhang}@xjtu.edu.cn, {zyb970305,405331375}@stu.xjtu.edu.cn).

signal processing methods; 3) data-driven methods.

The first type attempts to reproduce the saturated CT secondary current by modeling magnetization characteristics of the ferromagnetic core. Based on the measured magnetization curve [9]-[11], simplified magnetization curve [12], [13], and Fröhlich hysteresis technique [14], the magnetizing current could be accurately and efficiently estimated to correct the distorted CT secondary current. However, their effectiveness essentially depends on the precise estimation of the initial operating status of CT, which is difficult in practice.

The key of the signal processing methods is to distinguish the unsaturated waveforms from the saturated waveforms. To achieve this goal, the mutation information of the CT operating status transition is generally used. In [15], the section of the unsaturated CT current waveform was empirically determined, which may result in insufficient utilization of information and low accuracy. To automatically identify the unsaturated current waveform, multiple-order derivative information of the current waveform was used in [16]-[18]. Similarly, in [19], [20], the wavelet transform was used to extract the high-frequency component for the detection of the onset and endpoints of the CT saturation. However, the above methods are sensitive to measurement noises and dependent on the preset threshold, which limits their application potential. The extended Kalman filtering (EKF) [21] and morphological filtering techniques were used in [22], [23] to integrate the CT saturation detection and correction for its outstanding merits of accuracy and robustness against the noises. However, it is challenging to design proper filter parameters that suit different scenarios.

Data-driven methods, such as artificial neural networks (ANNs), have been introduced to cope with the problem of CT saturation [24], [25]. Although data-driven methods have powerful capabilities in data processing and feature mining, their interpretability and generalization are seriously concerned.

Each of the abovementioned methods has its pros and cons. Through the proper integration, it is greatly promising to realize the complementary advantages among the methods, and ultimately better performance. This hybrid-driven idea has been extensively explored in power systems [26]-[28] and other research fields [29], [30]. Inspired by these works, this paper proposes a hybrid physical model- and data-driven method to cope with the problem of CT saturation. The major contributions of this paper are listed as follows:

1) Tactful integration of physical model and data model is realized. The proposed hybrid-driven method is divided into two serial parts, namely the data-driven detection and physical model-driven compensation of CT saturation. The Fully Convolutional Network (FCN) [31] is built to detect the operation status of CT which is multifactorial and strongly nonlinear. Then a classic short-circuit current model is applied to reproduce the real current, and the model parameters are estimated by the efficient Levenberg-Marquardt nonlinear least squares (LM-NLS) algorithm [32]. The data-driven and physical model-driven methods deal with different parts of the CT saturation according to their respective advantages and finally solve this problem together.

2) The detection of CT saturation is more accurate and easier due to the powerful nonlinear modeling and feature extraction capabilities of FCN. Thus, the arduous selection of universal thresholds and model parameters in existing CT saturation detection methods is avoided. Moreover, FCN is computationally efficient and can process variable-length samples, which significantly enhances its applicability to various scenarios.

3) The generalization and interpretability of the proposed method are enhanced by introducing the short-circuit current model in the CT compensation part. These features provide a guarantee for the effectiveness of the method and greatly improve its application potential in future relay protection.

The rest of the paper is organized as follows: Section II presents the mechanism of CT saturation and the framework of the proposed method. In Section III, the implementation of the method is explained in detail. The effectiveness of the proposed method is validated by simulation results in Section IV. The experimental verification and its application potential are explored in Section V. Finally, Section VI draws the conclusion.

## II. MECHANISM OF CT SATURATION AND FRAMEWORK OF PROPOSED METHOD

In this section, the expression of the transient magnetic flux of CT under short-circuit current is derived under certain postulates, which reveals the mechanism and the major influence factors of CT saturation. Besides, the framework of the proposed hybrid-driven method is elaborated.

### A. Mechanism of CT saturation

As above mentioned, CT saturation is an intricate nonlinear and multifactorial problem. Although data-driven methods are independent of the physical models, a deep analysis of the mechanism of CT saturation is undoubtedly necessary to figure out the major influence factors, which conduces to construct a relatively complete sample database to improve the accuracy and generalization of the CT saturation detection model.

The equivalent circuit of CT converted to the secondary side is shown in Fig. 1 [9]-[11]. Concretely, $R$ and $L$ denote the resistance and inductance respectively, and subscripts 1, 2, $m$, and $l$ represent the primary side, secondary side, magnetizing branch, and load of CT.

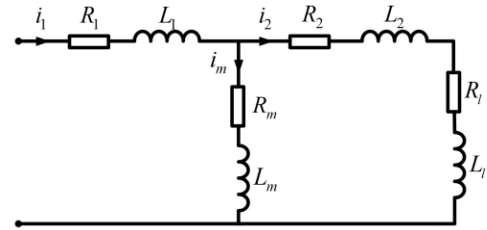

Fig. 1 The equivalent circuit of CT.

Normally, the CT secondary current reflects the system current accurately as the magnetizing current is negligible, as shown by the blue curve in Fig. 2. When a short-circuit fault occurs, for simplicity, it is assumed that the system operates in the no-load condition and the impedance angle of the short-circuit loop is $\pi/2$. Then, the short-circuit current can be expressed as (1) [6].

$$i_1(t) = I_m[\cos\theta e^{-t/T_1} - \cos(\omega t + \theta)], \quad (1)$$

where $I_m$ is the amplitude of the steady-state short-circuit current; $\theta$ is the initial phase angle of fault; $T_1$ is the system time constant; and $\omega = 2\pi f$, $f$ is the system frequency.

To analyze the transient characteristics of CT, the following simplifications and assumptions are essential.
1) Assuming the magnetizing inductance to be a constant.
2) Assuming the remanence to be zero.
3) Ignoring the magnetizing resistance.
4) Ignoring the leakage inductance and resistance of the primary side.

Based on the above assumptions, the following equation can be derived from the equivalent circuit shown in Fig. 1.

$$\left. \begin{array}{l} L_m \dfrac{di_m(t)}{dt} = L_s \dfrac{di_2(t)}{dt} + i_2(t)R_s \\ i_1(t) = i_2(t) + i_m(t) \end{array} \right\}, \quad (2)$$

where $L_s = L_2 + L_l$, $R_s = R_2 + R_l$.

Combining (1) and (2), the magnetizing current can be expressed as follows.

$$\begin{aligned} i_m(t) = &\dfrac{I_m}{\omega T_2}[-\sin(\omega t + \theta) + \sin\theta e^{-t/T_2} \\ &+ \dfrac{\omega T_1 T_2}{T_1 - T_2}\cos\theta(e^{-t/T_1} - e^{-t/T_2})] \end{aligned}, \quad (3)$$

where $T_2 = (L_m + L_s)/R_s$ represents the time constant of the secondary closed-loop circuit of CT.

Then the transient magnetic flux can be calculated by (4).

$$\begin{aligned} \phi(t) = &\dfrac{L_m i_m(t)}{N} + \phi_r = \dfrac{L_m I_m}{\omega N T_2}[-\sin(\omega t + \theta) + \\ &\sin\theta e^{-t/T_2} + \dfrac{\omega T_1 T_2}{T_1 - T_2}\cos\theta(e^{-t/T_1} - e^{-t/T_2})] + \phi_r \end{aligned}, \quad (4)$$

where $N$ is the number of turns of CT, and $\phi_r$ is remanence.

Based on (4), the transient magnetic flux under short-circuit current is as shown by the red line in Fig. 2. The comparison between the blue curve and red curve indicates that the magnetic flux under short-circuit current is much larger than that of the normal condition, which may cause CT to work in the nonlinear region, namely CT saturation.

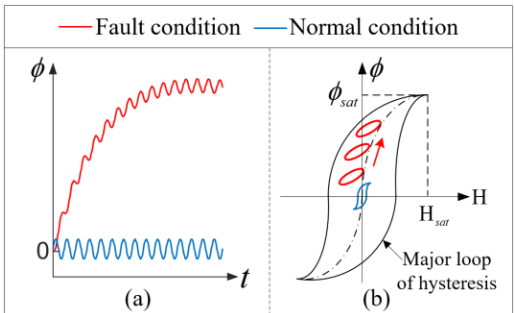

Fig. 2 Magnetic flux character of CT under fault condition and normal condition. (a) Transient magnetic flux curve. (b) Magnetization curve.

More importantly, the Eq. (4) reflects the major influence factors that associate with the transient magnetic flux of CT, which provides a theoretical basis for building a complete sample database by parameter traversal in the simulation model. Specifically, the major influence factors include the time constant of the system $T_1$, the initial phase angle of fault $\theta$, the amplitude of short-circuit current $I_m$, the time constant of the CT secondary closed-loop circuit $T_2$, the remanence $\phi_r$, and the nonlinearity of the magnetizing.

### B. Framework of the proposed method

The framework of the proposed method is shown in Fig. 3. Overall, the essential idea is to use the data-driven model to handle the complex and nonlinear characteristics of the CT in CT saturation detection and adopt a physical model, i.e. the short-circuit current model, to reproduce the real short-circuit current waveform. Hence, the tactful combination of the data model and the physical model is implemented in sequence.

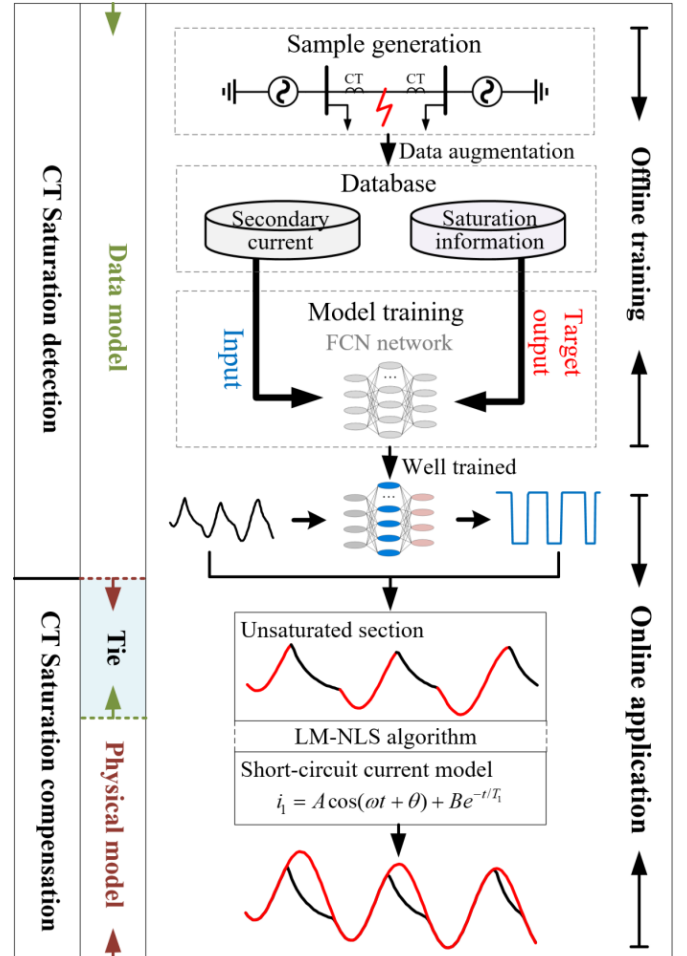

Fig. 3 Framework of the proposed method.

Concretely, saturation detection refers to monitoring the CT operation status and distinguishing the onset and endpoints of the saturated and unsaturated portions, while saturation compensation refers to restoring the saturated secondary current of CT. A complete sample database, the key to the accuracy and generalization of the data model, is established by traversing the major influence factors in the simulation model. Besides, the technology of data augmentation is applied to further improve the completeness of the database. The correlation between CT saturation and the influence factors is implicit in the sample database and then captured by the data model. The above process, including the database construction and the model training, is implemented offline. Then, the well-trained data model is deployed online to detect CT saturation.

As for the saturation compensation, the physical model of short-circuit current is retained for its conciseness and universality. Considering that the premises required by the (1) cannot be satisfied in engineering practice, a more accurate short-circuit current model is shown as follows.

$$i_1 = A\cos(\omega t + \theta) + Be^{-t/T_1} \quad (5)$$

Given the short-circuit current model in (5), saturation compensation is simplified as a problem of parameter estimation. After the saturated and unsaturated segments are distinguished by the data model, the saturated output waveform of CT can be efficiently compensated based on unsaturated data and the LM-NLS algorithm, as shown in Fig. 3. The framework of the proposed method is briefly introduced above, and the specific implementation will be elaborated in section III.

## III. IMPLEMENTATION OF PROPOSED METHOD

The details of the proposed method are explained here.

### A. The data-driven model for CT saturation detection

As shown in Fig. 3, the data-driven model consists of two parts, namely the construction of the sample database, the construction and training of the network. The sample database is constructed firstly based on the simulation model. Then, the technique of data augmentation is applied to improve the completeness of the sample database. To facility the application of the data model in postmortem analysis and online detection, a one-dimensional FCN is built which can efficiently process the variable-length samples.

*1) The construction of sample database*

As indicated in subsection II-A, CT saturation is mainly affected by the system parameters $T_1$, $\theta$, $I_m$, and CT parameters $T_2$, $\phi_r$. With the aid of power system simulation software, e.g., PSCAD/EMTDC, enormous simulation data can be generated through parameter traversal.

To train the FCN, the saturation information for samples is essential. Since the real short-circuit current and the CT secondary current can be expediently obtained from the simulation, the saturation information for each sample can be acquired by comparing the waveforms of primary current and secondary current which are denoted as $i_1$ and $i_2$, respectively. Thus, a straightforward criterion is given as follows.

$$S[k] = \begin{cases} 1, & \frac{|i_1[k] - i_2[k]|}{\max(|i_1|)} \geq \varsigma \\ 0, & else \end{cases}, k = 1, 2, \cdots n, \quad (6)$$

where $S$ denotes the serial saturation information and $n$ denotes the dimension of the sample. The threshold, $\varsigma$, determines the sensitivity of the saturation detection. Specifically, the smaller $\varsigma$ means higher sensitivity, but the robustness will decrease accordingly. Numerous simulations reveal that a reasonable value range is 0.01-0.1 and $\varsigma = 0.05$ is adopted in this paper. If CT is saturated at the $k$-th sample point, $S[k]$ is equal to 1; otherwise, it is equal to 0.

Moreover, to avert the problem of numerical instability, it is necessary to normalize the input of FCN, i.e., the secondary current of CT. The normalization is shown in (7).

$$i_2^*[k] = \frac{i_2[k]}{\max(\text{abs}(i_2))}, k = 1, 2, \cdots, n, \quad (7)$$

where $i_2^*$ denotes the normalized secondary current of CT with the value range of -1 to 1.

The normalized secondary current of CT and corresponding saturation information constitute a training sample, i.e. $(i_2^*, S)$, as shown in Fig. 4.

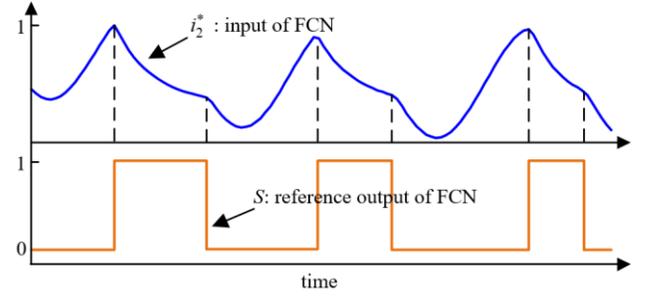

Fig. 4 Normalized secondary current and its corresponding saturation information.

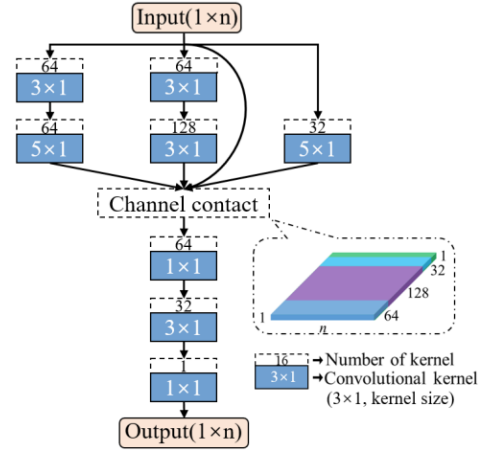

Fig. 5 Saturation detection model based on FCN.

*2) Data augmentation*

Data augmentation has been widely applied to improving the generalization capabilities of deep neural networks in recent years [33]. For the same purpose, the data augmentation is implemented in this paper, and details are given as follows.

● Polarity reversal

For each sample $(i_2^*, S)$, the polarity reversal is performed on the network input to form a new sample $(-i_2^*, S)$.

● Noise addition

For each sample $(i_2^*, S)$, a certain intensity Gaussian white noise is added in the network input to form a new sample $(noise(i_2^*, SNR), S)$, while $SNR$ represents the intensity of the noise. Considering that there are various low pass filters in the signal acquisition system, the noise intensity is generally small, so the value of the noise intensity is set as 40dB or 35dB.

● Variable-length window

For each sample $(i_2^*, S)$, intercepting a certain length can produce a new sample $(i_2^*[1:k], S[1:k]), k < n$. In this paper, the variable-length window can be 10ms, 20ms, 40ms, and 60ms, for coping with different application scenarios.

*3) The FCN for CT saturation detection*

Considering the limitations of the Convolutional Neural Network (CNN) and the Deep Belief Network on the input and output sizes, this paper uses the FCN to detect the CT saturation for its adaptability to the variation of input size. Moreover, although the Recurrent Neural Network (RNN) can also handle input signals of variable size, the sequential operation characteristic limits its computational efficiency. As a variant of the CNN, the FCN not only retains the computational efficiency of the CNN but has the ability of the RNN to process data with variable size. The specific structure of FCN for CT saturation detection is shown in Fig. 5. To maintain the consistency of input and output sizes, the network contains only convolution operations, and the convolution kernel is one-dimensional. Besides, for multiple receptive fields and fewer network parameters, the branch structure and unit convolution (i.e. 1*1 kernel) of GoogLeNet in [34] are used for reference when building the FCN as shown in Fig. 5.

### B. The physical model for CT saturation compensation

The essence of CT saturation compensation is to reproduce the real short-circuit current from the distorted secondary current of CT. Based on the physical model of short-circuit current in (5), the CT saturation compensation is essentially converted into a parameter estimation problem. In this paper, the LM-NLS algorithm is used to solve the above issue due to its advantage of stability and efficiency.

First, the short-circuit current $i_1$ is discretized according to the sampling interval.

$$i_1|_{t=k\cdot\Delta t} = i_1[k] = A\cos(\omega k\Delta t + \theta) + Be^{\lambda k\Delta t}, \quad (8)$$

where $\lambda = -1/T_1$ and $\Delta t$ denotes the sampling interval. $A$, $\theta$, $B$ and $\lambda$ are the parameters to be estimated.

The index sequence of the unsaturated sampling points identified by FCN is denoted as $I$.

$$I = [k_1 \quad k_2 \quad \cdots \quad k_N] \quad (9)$$

As CT is unsaturated, the sampling values of secondary current in the set $I$ are effective observed values of the short-circuit current. Therefore, saturation compensation is a parameter estimation problem based on a definite physical model and its effective observation data. The nonlinear least squares (NLS) algorithm is an efficient method for estimating the parameters of the nonlinear model as shown in (8). The specific implementation is given as follows.

The unsaturated data can form an error vector, $F$. Each vector element denotes the error between the observed value and the real value of the physical model at the sampling time.

$$F(x) = \begin{bmatrix} i_2[k_1] - A\cos(\omega k_1 \Delta t + \theta) + Be^{\lambda k_1 \Delta t} \\ \vdots \\ i_2[k_N] - A\cos[\omega k_N \Delta t + \theta] + Be^{\lambda k_N \Delta t} \end{bmatrix}, \quad (10)$$

where $x = [A \quad \theta \quad B \quad \lambda]$.

Hence, the parameter estimation problem in (5) is equivalent to the following least square problem.

$$x = \arg\min_x \frac{1}{2}\|F(x)\|^2 \quad (11)$$

The Jacobian matrix is essential to solve the least square problem in (11), as expressed in (12).

$$J = \begin{bmatrix} \cos(\omega k_1 \Delta t + \theta) & -A\sin(\omega k_1 \Delta t + \theta) & e^{\lambda k_1 \Delta t} & Bk_1\Delta t e^{\lambda k_1 \Delta t} \\ \vdots & \vdots & \vdots & \vdots \\ \cos(\omega k_N \Delta t + \theta) & -A\sin(\omega k_N \Delta t + \theta) & e^{\lambda k_N \Delta t} & Bk_N\Delta t e^{\lambda k_N \Delta t} \end{bmatrix} \quad (12)$$

To balance the stability and convergence speed, the Gauss-Newton method modified by the LM algorithm is applied in this paper, and the pseudo-code is given in TABLE I. In the LM-NLS algorithm, the damping parameter $\mu$ is introduced to deal with the numerical stability problem caused by the irreversibility of the Hessian matrix, $H$. The parameter $\rho$ is the gain ratio, which indicates the quality of $\mu$.

TABLE I
PSEUDO-CODE OF SATURATION COMPENSATION BASED ON LM-NLS

**Input:** $I = [k_1 \quad k_2 \quad \cdots \quad k_N]$; $i_2 = \{i_2[k_1] \quad i_2[k_2] \quad \cdots \quad i_2[k_N]\}$; initial parameters estimate $x_0 = [A_0 \quad \theta_0 \quad B_0 \quad \lambda_0]$.

**Output:** A vector $x^*$ minimizing $F(x)$ in (10).

**Algorithm:**
1. $k=0$, $x = x_0$, $\mu_0 = \tau * \max((J^T J)_{ii})$, $\varepsilon$
2. $H = J^T J$, $g = J^T F(x)$
3. if $\|g\|_2 < \varepsilon$, stop. $x^* = x$
4. $\Delta x = -(H + \mu_k I)^{-1} g$
5. $\rho = (\|F(x)\|^2 - \|F(x+\Delta x)\|^2)/(\Delta x^T(\mu_k \Delta x + g))$
6. if $\rho < 0.25$, $\mu_{k+1} = 2\mu_k$;
   else if $\rho > 0.75$, $\mu_{k+1} = \mu_k/2$; else, $\mu_{k+1} = \mu_k$
7. if $\rho > 0$, $x = x + \Delta x$
8. $k = k+1$, go to step 2

*Recommended value $\tau = 10^{-3}$, $\varepsilon = 10^{-15}$

## IV. SIMULATION VALIDATION

### A. Simulation model and network model

*1) Database construction by simulation*

A simulation model of a double-ended power system is built in PSCAD/EMTDC, as shown in Fig. 6. The system voltage is 220kV and the frequency is 50Hz. The turn ratio of CT is 2000:5. Note that the CT in Fig.6 adopts the JA model built-in PSCAD. The model can accurately simulate the magnetization process using domain wall motion with pinning effects to describe hysteresis loops of soft magnetic materials [35]. JA model is widely used due to the ease in numerical implementation and its fewer parameters.

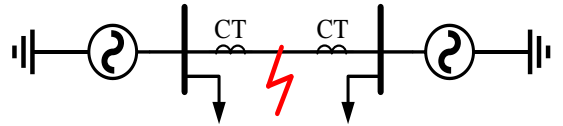

Fig.6 Schematic diagram of the simulation system.

TABLE II
VALUES OF MAJOR INFLUENCE PARAMETERS

| | Symbol | Quantity | Values |
|---|---|---|---|
| system | $T_1$ /ms | time constant | 50, 60, 70, …, 290, 300 |
| | $\theta$ | initial fault angle | 0°, 15°, …, 315°, 330° |
| | $R_f$ /Ω | fault resistance | 0, 1, 10, 50, 100 |
| CT | $T_2$ /s | time constant | 0.5, 0.75, 1, 1.25, 1.5, 1.75, 2 |
| | $\phi_r$ | residual flux | ±80%, ±40%, ±20%, 0 |

As discussed in section III-A, the database is constructed by the traversal of key parameters. Despite $I_m$ cannot be set directly, it can be indirectly controlled by the fault resistance, $R_f$. TABLE II presents the specific values of all major influence parameters. A total of 146,510 original samples have been obtained. For the balance of sample number, a part of unsaturated samples is removed and 102,571 valid samples have remained. Furthermore, the data augmentation technique in section III-A is used by three times, as a result of which the number of samples increases to 410,284.

*2) Network construction and training*

TensorFlow is one of the mainstream deep learning platforms developed by Google, which provides various high-level APIs to streamline the construction and training process of the network model. In this paper, TensorFlow is used for the construction and training of the saturation detection model.

The branched FCN network shown in Fig. 5 is constructed based on the Functional API while the one-dimensional convolutional layer is the built-in layer module of TensorFlow. Moreover, the activation function of the middle layers is *tanh*, and that of the output layer is *sigmoid*, so the network output is restrained in the range of 0-1.

In the supervised learning mode, the mean square error between the network output and the reference output is defined as the loss function, which forces the network output to gradually approach the reference output. Specifically, the Adam optimizer is adopted for gradient backpropagation and network parameter update. Except for the learning rate, the default parameter values of the Adam optimizer in TensorFlow are used (beta1=0.9, beta2=0.999, epsilon=1e-7). A descending learning rate is applied for better convergence performance, i.e., 1e-3 as the initial value and halved every 25 epochs. Besides, the training set and the test set samples are divided into 9:1.

*B. Effectiveness verification*

Firstly, the proposed hybrid-driven method is verified under different saturation scenarios, results shown in Fig.7. Fig. 7(a), (b), and (c) show that FCN can accurately distinguish between the saturated and unsaturated portion under different saturation levels, which verifies the high sensitivity of the data model for CT saturation detection. Fig. 7(b), (c), and (d) show that the start time of saturation, whether in the first cycle or non-first cycle, does not affect the accuracy of FCN, which extends the application of the proposed method. Furthermore, the effectiveness of the proposed saturation compensation model is demonstrated in Fig. 7 as the compensation result is almost the same as the real short circuit current of the power system. The excellent performance of the proposed compensation model is of great significance to prevent the relay protection devices from malfunction.

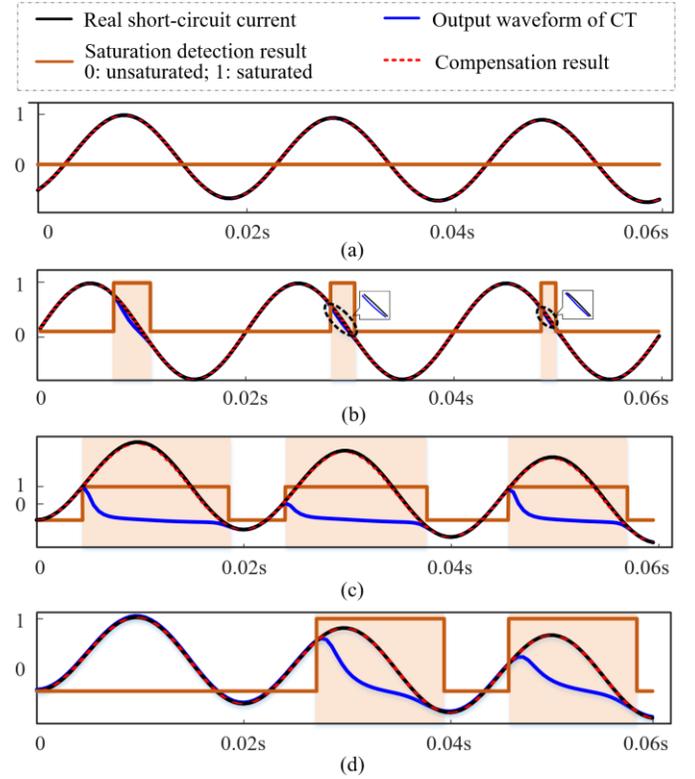

Fig. 7. Effect of proposed method on typical samples. (a) Unsaturated sample. (b) Slightly saturated sample. (c) Heavily saturated sample. (d) Non-first wave saturated sample.

TABLE III
COMPARISON OF PERFORMANCE INDEXES WITH/WITHOUT COMPENSATION

| | $e_1$ | $e_2$ | $e_3$ | $e_4$ |
|---|---|---|---|---|
| Without Compensation | 1.0001 | 0.7876 | 0.9693 | 0.5668 |
| Method in ref.[21] | 0.0736 | 0.0512 | 0.0754 | 0.0431 |
| Method in ref.[25] | 0.2640 | 0.1256 | 0.2740 | 0.1321 |
| Proposed Method | 0.0499 | 0.0241 | 0.0608 | 0.0153 |

To further verify the effectiveness of the proposed method, the performance indexes shown in (13) are proposed as follows.

$$\begin{cases} e_1 = \max_k \left\| i_1^{(k)} - i_2^{(k)} \right\|_\infty / \left\| i_1^{(k)} \right\|_\infty \\ e_2 = \frac{1}{N} \sum_{k=1}^N \left\| i_1^{(k)} - i_2^{(k)} \right\|_\infty / \left\| i_1^{(k)} \right\|_\infty \\ e_3 = \max_k \left( \left\| i_1^{(k)} - i_2^{(k)} \right\|_2 / \left\| i_1^{(k)} \right\|_2 \right) \\ e_4 = \frac{1}{N} \sum_{k=1}^N \left( \left\| i_1^{(k)} - i_2^{(k)} \right\|_2 / \left\| i_1^{(k)} \right\|_2 \right) \end{cases}, \quad k = 1, 2, \cdots, N \quad (13),$$

where $N$ denotes the number of samples. The indexes $e_1$ and $e_3$ represent the worst-case sample while $e_2$ and $e_4$ represent the overall performance. A smaller performance index value indicates that the CT measurement is more accurate.

TABLE III presents the comparison of performance indexes with/without the compensation. The EKF [21] and modified least square error [25] are selected as references to the proposed method. The results show that the performance indexes are reduced significantly after compensation, indicating that the above three methods can all effectively mitigate the CT saturation. Moreover, the proposed hybrid-driven method is more effective than the reference methods. After the compensation, the overall error decreases from 78% to 2.0% (2.41% and 1.53%) while the error of worst-case is no more than 6.1% (4.99% and 6.08%). Therefore, it is verified that the proposed method can effectively detect the saturation and accurately reproduce the real current waveform.

### C. Data window Flexibility and calculation efficiency

The proposed method can flexibly deal with variable-length samples due to the adoption of FCN. Fig. 8 shows the results of the proposed hybrid-driven method under different data window lengths, i.e. 0.01s, 0.06s, 0.02s, and 0.04s, respectively. The results indicate that the proposed method can accurately detect the saturated portion and compensate the current waveform as same as the real current waveform with variable data window length. This feature is of high engineering value, as the proposed method can be directly applied to various scenarios with different requirements for the data window without extra modification.

Moreover, the proposed method has high computational efficiency under different data window lengths. As the computation burden of the LM-NLS algorithm is much less than that of the FCN, the calculation time of the hybrid-driven method is dominated by the FCN. The relation between the computation time of the proposed method and the data window length is presented in Fig. 9, which is approximately linear as the time complexity of the one-dimensional FCN is exactly $O(N)$ if $N$ denotes the sample length. This linear characteristic is beneficial to highlight the flexibility and applicability of the proposed method. Specifically, with the hardware configuration shown in Fig. 9, the computation time is about 3ms in a considerable range of data windows, which can satisfy most scenarios in practice.

In summary, the proposed method is not only capable of dealing with variable-length samples, but also computationally efficient. This is a major aspect that the proposed method is superior to the existing methods based on ANN, which require a fixed window due to the network structure.

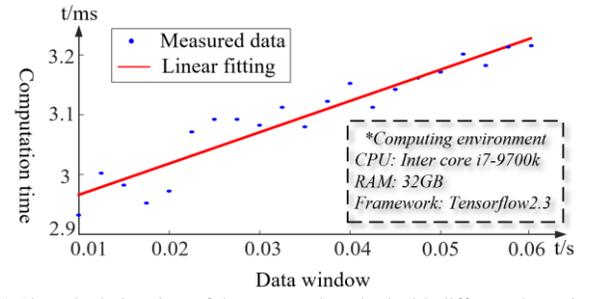

Fig. 9 The calculation time of the proposed method with different data window

### D. Sensitivity and robustness

In the issue of CT saturation, sensitivity emphasizes the ability to detect slight saturation while robustness represents anti-noise performance. The mutation information contained in the noise may be mistaken for CT saturation, which makes accurate saturation detection more difficult. Although various measures such as morphology filtering and Kalman filtering have been taken to improve the anti-noise ability of CT saturation detection methods, it remains a very challenging task to select a proper threshold to achieve the balance of sensitivity and robustness.

Fig. 10 shows the comparison of sensitivity and robustness between the proposed method and the state-of-the-art software-based method in [21]. The contrast between Fig. 14(a) and (b) indicates that a higher threshold, $\varepsilon$, can indeed improve the robustness of CT saturation detection, however at the cost of sensitivity. Conversely, reducing the threshold makes the method accurately recognize slightly saturated samples, but it weakens the anti-noise performance. Hence, for filter-based methods like [21], it is arduous to choose a proper threshold that can balance robustness and sensitivity. By contrast, the proposed hybrid-driven method can not only accurately detect light saturation samples, but also has good anti-noise performance, as shown in Fig. 10. The excellent balance of robustness and sensitivity is achieved by the proposed method as it avoids the complicated threshold parameters tuning.

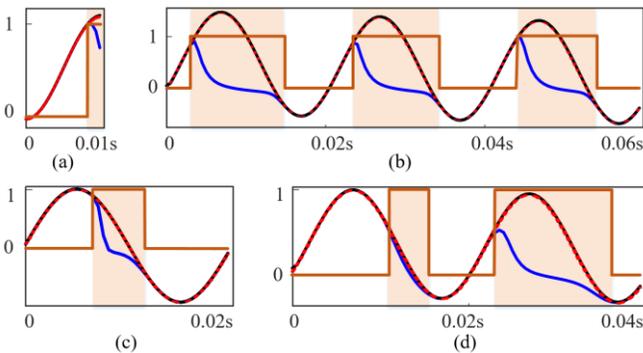

Fig. 8 Performance of the proposed method on samples with different data window length. (a) 0.01s. (b) 0.06s. (c) 0.02s. (d) 0.04s

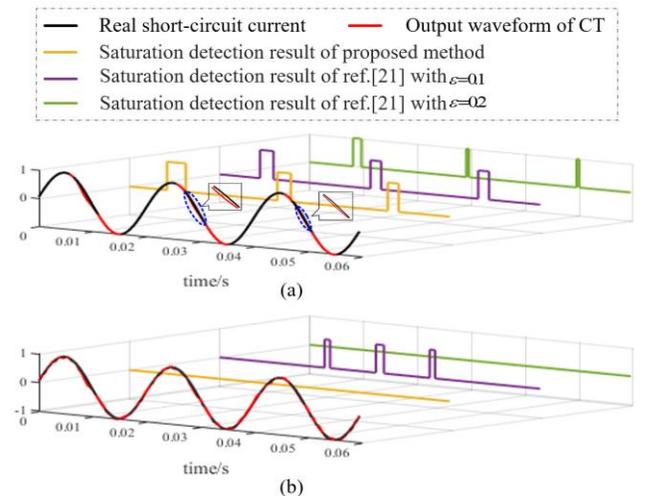

Fig. 10 Sensitivity and robustness comparison between the proposed method and reference [21] in (a) slightly saturated scenario without noise and (b) unsaturated scenario with 35dB noise.

## V. EXPERIMENTAL VALIDATION AND APPLICATION

Similar to other data-driven methods, the proposed hybrid-driven method faces the essential challenge that whether the network trained by certain CT model simulation data can be applied in scenarios with other CT models or even real CT devices. In this section, the hardware experiment is carried out to verify the generalization of the proposed method. Moreover, its potential application to relay protection is also explored.

### A. Hardware Experiment

Since the difference between CT models and real CT devices is inevitable, a hardware experiment is necessary to verify the validity of the proposed method in practice. In this regard, an experimental platform for CT saturation test was designed and established to obtain the real saturated current. The design and hardware platform are given in Fig. 11.

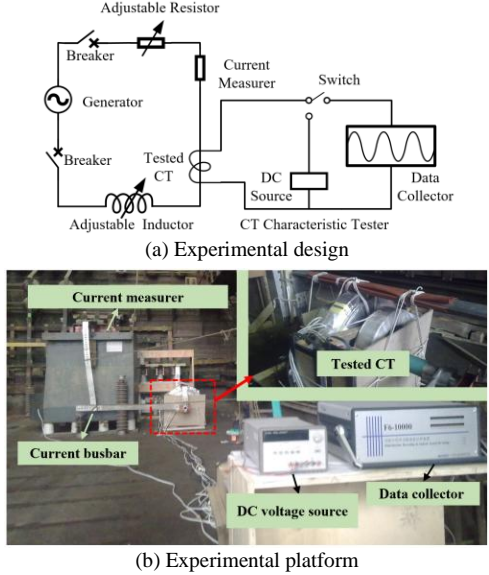

(a) Experimental design

(b) Experimental platform

Fig. 11. CT saturation hardware experiment

As Fig.11 shows, the primary circuit of the experiment consists of an ac generator, a current measurer, breaks, and the adjustable resistor and inductor. The primary current can be accurately measured by a high-power precise non-inductive resistance, i.e. the current measurer. The adjustable resistor and inductor are used to realize different short-circuit currents. The secondary current is measured by the tested CT devices via the current busbar. The DC voltage source is equipped to demagnetize CT if necessary.

Note that the short-circuit current is synthesized by Fig.11. The damping dc component of the short-circuit current is changed by the breaker with a different closing angle. The adjustable resistor dominates the time-constant of dc component while the adjustable inductor controls the periodic ac component of the short-circuit current. By changing the inductance, resistance, and closing angle of the primary circuit, different short-circuit currents can be synthesized to adequately test the transient transformation characteristics of CT.

Three types of CT devices are tested to fully verify the performance of the proposed method. The details of the tested CTs are listed in TABLE IV. 'SH' and 'ZX' represent different manufacturers while '5P20' and '5P30' indicate the accuracy level. For example, '5P20' requests that the composite error of secondary current is less than ±5% when the primary current is 20 times the rated current.

TABLE IV
PARAMETERS OF THE EXPERIMENTAL CT

| CT type | Accuracy level | Transformation ratio | Rated capacity | Power factor |
|---|---|---|---|---|
| SH5P20 | 5P20 | 1:1000 | 20VA | 1 |
| SH5P30 | 5P30 | 1:1000 | 20VA | 1 |
| ZX5P20 | 5P20 | 1:1000 | 20VA | 0.8 |

For each type of CT device, 20 sets of experiments were performed, and a total of 60 sets of experimental data were obtained. For better verification, the experimental results include the unsaturated, slightly saturated, and severely saturated current waveforms.

### B. Generalization performance verification

In this section, the generalization verification of the proposed method is twofold. On the one hand, the JA model and the Lucas model [36] are adopted for simulation verification. Different from the JA model, the Lucas model is based on the technical scheme of establishing the equivalent nonlinear circuit of CT. According to the simulation system shown in Fig. 6, the Lucas CT model built-in PSCAD is used to generate the data set, the parameter traversal shown in TABLE II. On the other hand, a more realistic test is carried out with the real CT measurement data obtained by the above hardware experiment.

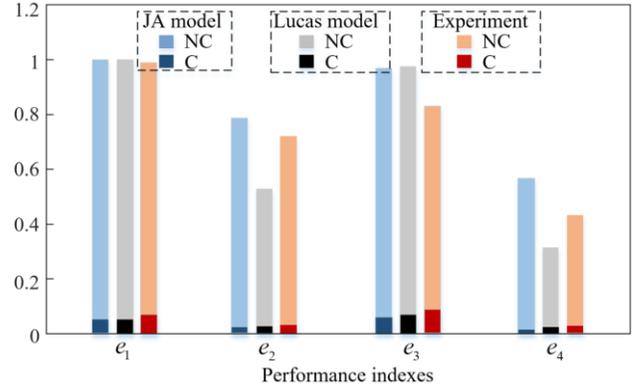

Fig. 12 Comparison of performance indexes with/without the proposed method for different CT model/experiment data set.

Fig. 12 shows the performance of the proposed method in the JA model data set (410,284 samples), Lucas model data set (146,510 samples), and experimental data set (60 samples). The indicators "C" and "NC" in Fig. 12 represent the samples with and without the proposed saturation compensation method, respectively. The performance indexes without compensation are large, indicating that the current saturation is considerable in three data sets. However, the performance indexes are reduced significantly after the compensation, which verifies that the saturated currents are effectively compensated with the proposed method in all data sets. Though the performance indexes in the Lucas model data set and experiment data set are slightly larger than that in the JA model data set, the overall error ($e_2$ and $e_4$) is less than 2.85% (2.76% and 1.94% for Lucas model data set, 2.79% and 2.85% for experiment data set). In conclusion, though trained by JA model simulation data, the proposed method is accurate and effective in the scenarios with other CT models or CT devices.

## C. Application Potential in relay protection

To explore the application potential, the proposed method is assumed to integrate into the relay protection and verify whether the performance of the protection is improved. In this regard, the current differential protection principle, one classical protection principle, is adopted.

Denote $I_{op}$ as the operating current and $I_{re}$ as the restraining current, which can be expressed as follows [37]:

$$I_{op} = |\dot{I}_{ij} + \dot{I}_{ji}|, I_{re} = |\dot{I}_{ij} - \dot{I}_{ji}| \quad (14)$$

The percentage differential relay operates when

$$I_{op} \geq I_o \,\&\, I_{op} \geq KI_{re}, \quad (15)$$

where $I_o$ is a pick-up current and $K \in (0,1)$ is the restraint coefficient, and K=0.3 is adopted in this paper. The data window of the differential protection is 20ms.

The differential protection is equipped on a transmission line $i$-$j$. Assume that the CT on the $i$-th side is never saturated while the CT on the $j$-th side is one of the tested CTs. Under the condition of external fault, $\dot{I}_{ij}$ is the same with primary current provided by the current measurer and $\dot{I}_{ji}$ is measured by the tested CTs provided by the data collector in Fig. 11.

TABLE V compares the malfunction rate of the protection with/without the proposed method. The results show that the overall malfunction rate is 23/60 before the compensation, indicating that the protection performance is seriously influenced by the CT saturation. After the compensation by the proposed method, all malfunctions are effectively avoided, and the protection performance is guaranteed.

TABLE V
PERFORMANCE OF PROPOSED METHOD IN ANTI-MALFUNCTION OF CURRENT DIFFERENTIAL PROTECTION

| CT type | Malfunction rate | |
|---|---|---|
| | without proposed method | with proposed method |
| SH5P30 | 6/20 | 0/20 |
| SH5P20 | 7/20 | 0/20 |
| ZH5P20 | 10/20 | 0/20 |

Fig. 13 shows how the proposed method avoids the malfunction of current differential protection that is caused by CT saturation. The measured current waveforms, operating current and restraining current, and the output signal of current differential protection are given in Fig.13. Fig. 13(a), (b), and (d) show that the saturated current produces a considerable operating current and finally leads the protection to malfunction. After compensation by the proposed method in Fig. 13(a), the saturated current is accurately detected and properly compensated, thus the operating current is quietly small in Fig. 13(c). The malfunction is eventually avoided as shown in Fig.13(d).

## D. Discussion on time delay

If the saturation compensation method is integrated into the protection, its calculation time may cause a time delay. Fig.14 illustrates the influence of such time delay on relay operation.

As Fig. 14 shows, for the scenarios without saturation compensation, the primary protection will trigger immediately after the fault is detected. Denote $\tau_1$ as the length of the data window and $\tau_2$ as the calculation time of the protection, then the trigger time of the primary protection is $t_p = \tau_1 + \tau_2$. For the backup protection with a time delay $T_{delay}$, the trigger time is $t_b = t_p + T_{delay}$. Note that the relay is continuously calculating with the sliding data window during the delay time. Generally speaking, $\tau_2$ is larger than the sampling time interval, thus the updating period of the sliding data window is $\tau_2$, as shown in Fig.14.

For the scenarios with the proposed method, the primary protection operates after obtaining the data window, compensating the saturated currents, and calculating the protection. Thus, the trigger time of the primary protection is $t'_p = \tau_1 + \tau_3 + \tau_2$, which leads to a time delay of $\tau_3$ compared with the scenarios without compensation. Similarly, the trigger time of the backup protection with the proposed method is $t'_b = t'_p + T_{delay}$, and a time delay of $\tau_3$ is inevitable. In conclusion, a time delay of $\tau_3$ is introduced for the relay protection after the integration of the proposed method. Compared with the benefits of avoiding protection malfunctions, the slight time delay by the integration of the proposed method is worthwhile.

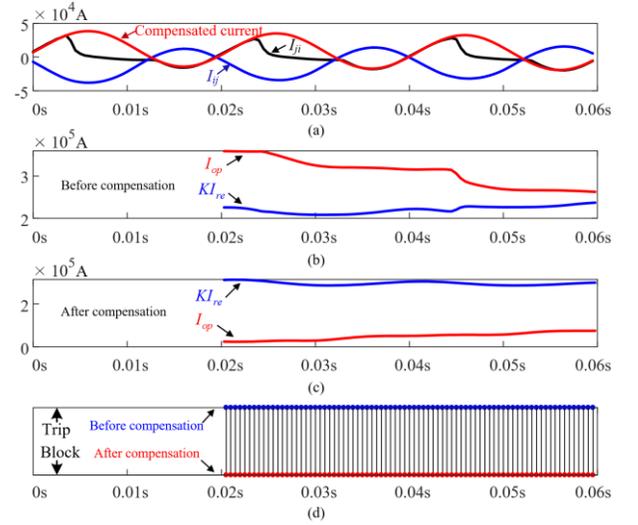

Fig. 13. Protection-related sequential information. (a) Measured current waveform. (b) Operating current and restraining current before saturation compensation. (c) Operating current and restraining after saturation compensation. (d) Trip or block signal of current differential protection.

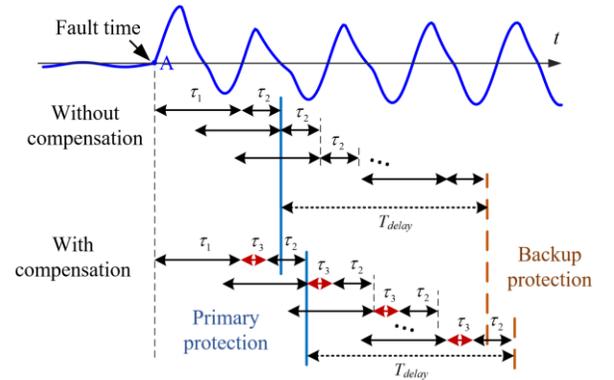

Fig. 14 The time-delay caused by the proposed method in relay protection.
$\tau_1$: length of data window, $\tau_2$: updating period of data window
$\tau_3$: time delay of the proposed method, $T_{delay}$: delay time of backup protection

It should be noted that the updating period of the sliding data window changes to $\tau_2+\tau_3$ with the proposed method. In other words, the updating period of the data window is extended by $\tau_3$. This restriction is acceptable for the protection as the general value for $\tau_2$ is 3-5ms while the calculation time of the proposed method is nearly 3ms. Moreover, the calculation time can be further reduced with the development of chip technology. It has been reported that the calculation speed of Google's AI chip TPU is 15-30 times faster than that of GPU/CPU [38]. Therefore, though the proposed method could not be deployed in actual protection devices presently due to the limited computing power of the microprocessor, it has great application potential in future protection.

## VI. Conclusion

In this paper, a hybrid physical model- and data-driven method is proposed for CT saturation detection and compensation. A one-dimensional FCN is constructed to detect CT saturation and the physical model of short-circuit current is used to compensate saturated current. The major merits of the proposed method consist in the tactful integration of the physical model and data model. The special structure of FCN endows it with the flexibility of processing variable-length samples, which is the main aspect that the proposed method is superior to the existing ANNs-based methods. Compared with the state-of-the-art software-based methods, the proposed method can effectively balance robustness and sensitivity without arduous parameters tuning. The hardware experiments further verify the generalization of the proposed method and explore its application potential in relay protection. The malfunctions of the protections caused by CT saturation can be effectively avoided by the proposed method at the cost of a slight time delay.

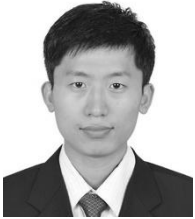

**Songhao Yang.** (S'18-M'19) was born in Shandong, China, in 1989. He received the B.S. and Ph.D. degrees in electrical engineering from the Xi'an Jiaotong University, Xi'an, China, in 2012 and 2019, respectively. Besides, he received the Ph.D. degree in electrical and electronic engineering from Tokushima University, Japan, in 2019.

Currently, he is an Assistant Professor at Xi'an Jiaotong University. His research interest includes power system control and protection.

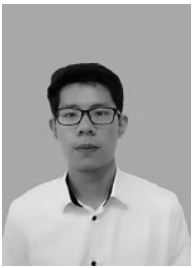

**Yubo Zhang.** (S'21) received the B.S. degree in electrical engineering from Xi'an Jiaotong University, China, in 2019. He is currently pursuing the Ph.D. degree in the school of Electrical Engineering in Xi'an Jiaotong University. His main field of interest includes the control for renewable energy, and power system frequency stability.

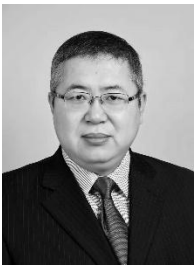

**Zhiguo Hao.** (M'10) was born in Ordos, China, in 1976. He received his B.Sc. and Ph.D. degrees in electrical engineering from Xi'an Jiaotong University, Xi'an, China, in 1998 and 2007, respectively. He has been a Professor with the Electrical Engineering Department, Xi'an Jiaotong University. His research interest includes power system protection and control.

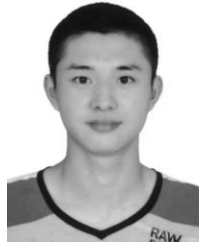

**Zexuan Lin.** received the B.S. degree from Shandong University, Jinan, China, in 2016. He is currently working toward M.S. degree in Xi'an Jiaotong University, Xi'an, China. His research intersests is the application of artificial intelligence in the relay protection.

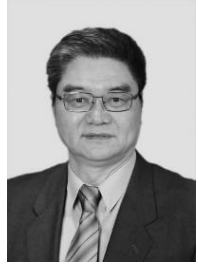

**Baohui Zhang.** (SM'99-'F'19) was born in Hebei Province, China, in 1953. He received the M.Eng. and Ph.D. degrees in electrical engineering from Xi'an Jiaotong University, Xi'an, China, in 1982 and 1988, respectively. He has been a Professor in the Electrical Engineering Department at Xi'an Jiaotong University since 1992. His research interests are system analysis, control, communication, and protection.